\documentclass[12pt]{article}

\usepackage{arxiv}

\usepackage[utf8]{inputenc} 
\usepackage[T1]{fontenc}    
\usepackage{hyperref}       
\usepackage{url}            
\usepackage{booktabs}       
\usepackage{amsfonts}       
\usepackage{amsmath} 
\usepackage{nicefrac}       
\usepackage{microtype}      
\usepackage{lipsum}
\usepackage{graphicx}
\usepackage{threeparttable}
\usepackage{multirow}
\graphicspath{ {./images/} }
\usepackage{setspace}
\title{Prevalence Estimation in Infectious Diseases with Imperfect Tests: A Comparison of Frequentist and Bayesian Logistic Regression Methods with Misclassification Correction}

\author{
 Jorge Mario Estrada Alvarez \\
  Caja de Compensacion Familiar de Risaralda\\
  Salud - Comfamiliar\\
  Pereira, PA 66003 \\
  \texttt{jestradaa@comfamiliar.com} \\
   \And
 Henan F. Garcia \\
  SISTEMIC Research Group\\
  Faculty of engineering\\
  University of Antioquia\\
  Medellin, PA 15213 \\
  \texttt{hernanf.garcia@udea.edu.co} \\
  \And
 Miguel Ángel Montero-Alonso \\
  Department of Statistics and Operational Research\\
  University of Granada\\
  Granada, Spain PA 18071 \\
  \texttt{mmontero@ugr.edu} \\
  \And
 Juan de Dios Luna del Castillo \\
  Department of Statistics and Operational Research\\
  University of Granada\\
  Granada, Spain PA 18071 \\
  \texttt{jdluna@ugr.edu} \\
}

\begin{document}
\maketitle
\begin{abstract}
Accurate estimation of disease prevalence is critical for informing public health strategies. Imperfect diagnostic tests can lead to misclassification errors, such as false positives and false negatives, which may skew estimates if not properly addressed. This study compared four statistical methods for estimating the prevalence of sexually transmitted infections (STIs) and the factors associated with them, while incorporating corrections for misclassification. The methods examined were: (i) standard logistic regression with external correction using known sensitivity and specificity; (ii) the Liu et al. model, which jointly estimates false positive and false negative rates; (iii) Bayesian logistic regression with external correction; and (iv) a Bayesian model with internal correction that utilizes informative priors on diagnostic accuracy.
Data were obtained from 11,452 individuals participating in a voluntary screening campaign for HIV, syphilis, and hepatitis B from 2020 to 2024. Prevalence estimates and regression coefficients were compared across models using metrics of relative change from crude estimates, confidence interval width, and coefficient variability.
The Liu model yielded higher prevalence estimates but exhibited wider intervals and convergence issues for low-prevalence conditions. In contrast, the Bayesian model with internal correction (BEC) provided intermediate estimates with the narrowest confidence intervals and more stable intercepts, reflecting an improved estimation of baseline prevalence. Prior distributions either informative or weakly informative contributed to regularization, particularly in small-sample or rare-event settings.
Furthermore, accounting for misclassification impacts both prevalence estimates and covariate associations. Although the Liu model presents theoretical advantages compared to standard regression, its practical limitations, particularly in cases with sparse data, diminish its effectiveness. In contrast, Bayesian models that incorporate misclassification correction provide a robust and flexible alternative in low-prevalence situations, proving especially useful when accurate estimates are necessary despite diagnostic uncertainties.
\end{abstract}

\keywords{Prevalence estimation \and Missclassification \and Logistic regression, Infectious disease}

\section{Introduction}

Accurate estimation of the prevalence of infectious diseases is essential for guiding the planning, implementation, and evaluation of public health strategies. Such estimates are typically based on the large-scale application of diagnostic tests, whose performance is rarely perfect. As a result, misclassification errors such as false positives and false negatives occur, introducing bias and reducing the precision of the resulting estimates \cite{Magder}.

Accounting for possible misclassification in the binary outcome variable is crucial, as ignoring it can lead to substantial bias in parameter estimates. Suppose misclassification is suspected, and information on the misclassification rates (sensitivity and specificity) is available or can be reasonably assumed. In that case, methods that explicitly incorporate this information should be employed to correct the estimated prevalence and avoid bias toward the null value in the resulting estimates \cite{Magder}.

Logistic regression is a commonly used tool for prevalence estimation that allows for covariate adjustment. Moreover, correcting for measurement error in logistic regression models improves the estimation of the prevalence of a condition while allowing the direct inclusion of relevant demographic and epidemiological covariates \cite{Fraser}. Nonetheless, methods that integrate both elements are required to adequately address misclassification's challenges and benefit from the flexibility to include covariates of interest. This combination enables the derivation of prevalence estimates with greater validity, particularly when multiple infectious diseases are epidemiologically correlated.

Various strategies have been developed to handle diagnostic errors in prevalence estimation, ranging from post-hoc adjustments via marginal standardization of predicted probabilities \cite{Muller}, which provides an extrinsic adjustment based on the model's predictions. In this same line of work, Liu et al. \cite{Liu} propose a logistic regression model with correction for misclassification in the binary outcome variable. The method incorporates parameters for the false positive (FP) and false negative (FN) rates into the standard logistic regression model, thereby enabling the estimation of coefficients adjusted for these error rates. The model proposed by Liu et al. includes different variants depending on whether one or both types of error need to be estimated, assuming these errors may be either equal or distinct.

Valle et al. \cite{valle} propose a Bayesian inference approach to logistic regression for correcting bias in prevalence estimates when using imperfect tests. However, their results primarily focus on the bias induced in the estimation of risk factor associations rather than on the method for adjusting prevalence estimates in the presence of covariates.

This study directly addresses the prevalence estimation of diseases using imperfect tests while accounting for the need to control for epidemiological and/or demographic covariates. The objective was to compare four statistical methodologies for prevalence estimation under a practical framework for adjusted and corrected point estimates, emphasizing the importance of an integrated approach that considers diagnostic uncertainty and covariate adjustment. The proposed methodological approaches (frequentist and Bayesian) were: (i) standard logistic regression for prevalence estimation adjusted for covariates, with correction based on known test sensitivity and specificity, (ii) modified logistic regression for simultaneous estimation of test errors as proposed by Liu et al. \cite{Liu}, (iii) standard logistic regression with Bayesian inference and extrinsic correction using known sensitivity and specificity, and (iv) a Bayesian logistic regression model with intrinsic correction of misclassification error via prior distribution elicitation for classification errors \cite{valle}. The methods are compared using a representative sample of 11452 individuals simultaneously evaluated for HIV and syphilis seroprevalence.

\section{Methods}

This study aims to estimate the prevalence of low-prevalence sexually transmitted infections (STIs) while correcting for diagnostic misclassification. We implemented and compared four logistic regression models that differ in their handling of misclassification and statistical inference frameworks.

\subsection{Crude Prevalence Estimation}

Crude prevalence was estimated using the Rogan--Gladen method \cite{rogan}, which adjusts the observed proportion of positive tests based on known sensitivity $(Se)$ and specificity $(Sp)$ of the diagnostic test:
\begin{align}
\hat{P}_{\text{adj}} = \frac{\hat{P}_{\text{obs}} + Sp - 1}{Se + Sp - 1}.
\end{align}
This estimator assumes fixed, known values of Se and Sp and yields an unbiased prevalence estimate under nondifferential misclassification.

\subsection{Adjusted Prevalence Estimation via Regression Models}

Four logistic regression models were applied to estimate adjusted prevalence for HIV and syphilis, accounting for both covariates and test misclassification. Covariates included age, sex, test results for other infections, and key population indicators. Table~\ref{tab:coeficientes} defines the regression coefficients associated with each covariate.

\begin{table}[ht]
\centering
\caption{Coefficient assignment for covariates in the models}
\label{tab:coeficientes}
\begin{tabular}{lc}
\toprule
\textbf{Covariate} & \textbf{Assigned Coefficient} \\
\midrule
Intercept & $\beta_0$ \\
Age (years) & $\beta_1$ \\
Sex & $\beta_2$ \\
Syphilis test* & $\beta_3$ \\
Hepatitis B test & $\beta_4$ \\
MSM** & $\beta_5$ \\
LGTBI population & $\beta_6$ \\
Other populations & $\beta_7$ \\
Sex worker & $\beta_8$ \\
\bottomrule
\end{tabular}
\begin{tablenotes}
\small
\item \textit{*} The coefficient $\beta_3$ for the syphilis prevalence models represents the result of the HIV test.
\item \textit{**} MSM: Men who have sex with men.
\end{tablenotes}
\end{table}

\subsubsection{Standard Logistic Regression (STD)}

The standard model estimates the probability of disease given covariates using the logistic function:
\begin{align}
P(Y_i = 1 \mid X_i) &= \pi(X_i) = \frac{\exp(\eta_i)}{1 + \exp(\eta_i)}, \\
\eta_i &= \beta_0 + \sum_{j=1}^p \beta_j X_{ij}.
\end{align}
Parameters are estimated via maximum likelihood. Adjusted prevalence is computed by correcting the predicted values using the Rogan--Gladen formula.

\subsubsection{Bayesian Logistic Regression with External Correction (BC)}

In the Bayesian extension, regression coefficients are assigned prior distributions. Following Newman et al. \cite{newman}, priors are:
\begin{align}
\beta_j \sim \mathcal{N}\left(0, \frac{\pi^2}{3(p+1)}\right), \quad j = 0, \dots, p.
\end{align}
Posterior inference is conducted via Markov chain Monte Carlo (MCMC). Point estimates are then corrected post hoc using known Se and Sp.

\subsubsection{Logistic Regression with Simultaneous Estimation of Misclassification (Liu)}

The Liu model \cite{Liu} introduces latent variables $\tilde{Y}_i$ for true disease status, linking them to observed outcomes $Y_i$ via misclassification probabilities, following traditional assumptions on non-differential misclassification \cite{Neuhaus,Hausman}:
\begin{align}
P(Y_i = 1 \mid \tilde{Y}_i = 0) &= r_0, \\
P(Y_i = 0 \mid \tilde{Y}_i = 1) &= r_1.
\end{align}
The conditional probability is then:
\begin{align}
P(Y_i = 1 \mid X_i) = r_0 + (1 - r_0 - r_1) \cdot \frac{1}{1 + \exp(-\eta_i)}.
\end{align}
Model parameters $(\beta, r_0, r_1)$ are jointly estimated via maximum likelihood.

\subsubsection{Bayesian Logistic Regression with Internal Correction (BEC)}

This hierarchical model \cite{valle} assumes the true outcome $\tilde{Y}_i$ follows a logistic model:
\begin{align}
\tilde{Y}_i &\sim \text{Bernoulli}(\pi_i), \\
\pi_i &= \frac{\exp(\eta_i)}{1 + \exp(\eta_i)}.
\end{align}
The observed test result $Y_i$ depends on $\tilde{Y}_i$ and the test characteristics:
\begin{align}
P(Y_i = 1 \mid \tilde{Y}_i = 1) &= Se, \\
P(Y_i = 1 \mid \tilde{Y}_i = 0) &= 1 - Sp.
\end{align}
Priors are specified as:
\begin{equation}
    Y_i \sim \text{Bernoulli}(Se) \quad \text{if } \tilde Y_i = 1,
\end{equation}

\begin{equation}
    Y_i \sim \text{Bernoulli}(1 - Sp) \quad \text{if } \tilde Y_i = 0,
\end{equation}
Inference is conducted via MCMC to obtain posterior distributions of prevalence and covariate effects.

\subsection{Evaluation Criteria}

Models were compared using the following metrics, according to the approach suggested by Morris et al., \cite{Morris}:

\begin{itemize}
  \item \textbf{Relative change in adjusted prevalence} with respect to crude estimates.
  \item \textbf{Interval width} of confidence/credible intervals.
  \item \textbf{Stability of regression coefficients}, measured by standard errors or posterior variance.
\end{itemize}

\subsection{Data Source and Variables}

Data were obtained from 11,452 individuals screened between 2020--2024. Covariates included age, sex, syphilis and hepatitis B status, and population group classification (e.g., MSM, sex worker). Tests used were the Determine HIV Early Detect (Se = 0.975, Sp = 0.999) and Bioline Syphilis 3.0 (Se = 0.964, Sp = 0.974).

\subsection{Software}

All analyses were performed in \texttt{R} version 4.4.3 \cite{rsoftware}. Frequentist models used base \texttt{glm()} and \texttt{optim()} routines; Bayesian models were estimated using \texttt{rstan} and \texttt{rstanarm} \cite{rstan, rstanarm}.

\section{Results}

\subsection{Sample Characteristics}

The study population included 11452 individuals, whose demographic and serological characteristics are summarized in Table~\ref{tab:descriptives}. The mean age was 34 years (standard deviation: 14). The sex distribution was balanced, with 5759 men corresponding to 50\%  of the sample.

Regarding serological results, 155 participants (1.4\%) tested reactive for HIV, 905 (7.9\%) for syphilis, and 8 (<0.1\%) for hepatitis B.

Concerning key population categories, the most significant proportion corresponded to the general population, with 6574 individuals (57\%), followed by men who have sex with men (MSM), with 3248 (28\%), and sex workers, with 1213 (11\%). Additionally, individuals identified as part of the LGTBIQ community (n = 224; 2.0\%) and other populations (n = 193; 1.7\%) were included.

\begin{table}
\centering
\caption{Descriptive characteristics of the study population (n = 11,452)}
\label{tab:descriptives}
\begin{tabular}{lc}
\toprule
\textbf{Characteristic} & \textbf{ } \\
\midrule
Age (mean [SD]) & 34 (14) \\
Male sex & 5759 (50\%) \\
HIV reactivity & 155 (1.4\%) \\
Syphilis reactivity & 905 (7.9\%) \\
Hepatitis B reactivity & 8 (<0.1\%) \\
\\[-1ex]
\multicolumn{2}{l}{\textbf{Key population type}} \\
\hspace{1em} MSM & 3248 (28\%) \\
\hspace{1em} LGTBIQ & 224 (2.0\%) \\
\hspace{1em} Other populations & 193 (1.7\%) \\
\hspace{1em} General population & 6574 (57\%) \\
\hspace{1em} Sex worker & 1213 (11\%) \\
\bottomrule
\end{tabular}
\begin{tablenotes}
\small
\item \textit{ } Continuous variables are presented as mean (standard deviation), and categorical variables as n (percentage).
\end{tablenotes}
\end{table}

\subsection{Comparison of Adjusted Prevalence Estimates}

Table~\ref{tab:prevalencia_vih_sifilis} presents the adjusted prevalence estimates for HIV and syphilis under different regression models with misclassification correction. Notable variations in the estimates were observed depending on the methodological approach used.

For HIV, the crude prevalence was 1.39\%. The standard model (STD), which applies an external correction based on known sensitivity and specificity, reduced this estimate to 0.73\%. In contrast, the Liu model, which simultaneously estimates diagnostic error parameters, yielded a prevalence of 1.66\%, representing a relative increase of 127.69\% compared to the STD model. Meanwhile, the Bayesian model with misclassification correction (BEC) estimated a prevalence of 1.00\%, with a confidence interval width of 0.0049.

In the case of syphilis, the crude prevalence was 5.67\%. The STD model produced a considerably lower adjusted estimate (2.77\%), while the Liu model estimated a prevalence of 11.55\%—more than twice the crude value and with the widest confidence interval (0.0187). In contrast, the BEC model estimated a prevalence of 4.14\%, with the narrowest confidence interval (0.0073), indicating the highest relative precision among the models compared.

Overall, the Bayesian models demonstrated a better balance between misclassification error adjustment and estimate precision—particularly the BEC model, which yielded intermediate prevalence values and the narrowest confidence intervals for both outcomes.

\begin{table}
\centering
\caption{Adjusted prevalence estimates for HIV and syphilis according to different regression models}
\label{tab:prevalencia_vih_sifilis}
\resizebox{0.95\linewidth}{!}{\begin{tabular}{llcccccc}
\toprule
\textbf{Disease} & \textbf{Model} & \textbf{Adjusted P} & \textbf{Lower} & \textbf{Upper} & \textbf{Change vs. Crude (\%)} & \textbf{Change vs. STD (\%)} & \textbf{CI Width} \\
\midrule
\multirow{5}{*}{HIV} 
  & Crude & 0.0139 & 0.0117 & 0.0160 & -- & -- & 0.0043 \\
  & STD   & 0.0073 & 0.0055 & 0.0097 & -47.42 & -- & 0.0042 \\
  & Liu   & 0.0166 & 0.0131 & 0.0212 & 19.72  & 127.69 & 0.0081 \\
  & BC    & 0.0095 & 0.0078 & 0.0116 & -31.57 & 30.15  & 0.0037 \\
  & BEC   & 0.0100 & 0.0078 & 0.0127 & -28.24 & 36.47  & 0.0049 \\
\midrule
\multirow{5}{*}{Syphilis}
  & Crude & 0.0567 & 0.0514 & 0.0620 & -- & -- & 0.0106 \\
  & STD   & 0.0277 & 0.0223 & 0.0336 & -51.21 & -- & 0.0113 \\
  & Liu   & 0.1155 & 0.1064 & 0.1252 & 103.64 & 317.39 & 0.0187 \\
  & BC    & 0.0340 & 0.0288 & 0.0391 & -40.08 & 22.82  & 0.0103 \\
  & BEC   & 0.0414 & 0.0379 & 0.0452 & -26.94 & 49.74  & 0.0073 \\
\bottomrule
\end{tabular}}
\begin{tablenotes}
\small
\item \textit{Note:} STD = standard model with external correction; Liu = model with simultaneous estimation of error parameters; BC = Bayesian model without internal correction; BEC = Bayesian model with misclassification correction.
\end{tablenotes}
\end{table}

\begin{table}
\centering
\caption{Comparison of regression coefficients for HIV according to the implemented model}
\label{tab:coeficientes_vih}
\begin{tabular}{lcccccc}
\toprule
\textbf{Coefficient} & \textbf{STD} & \textbf{Liu} & \textbf{Liu Change (\%)} & \textbf{BC} & \textbf{BEC} & \textbf{BEC Change (\%)} \\
\midrule
$\beta_0$                         & -5.864 & -4.648 & -20.73  & -5.418 & -4.717 & -12.93 \\
$\beta_1$                         & -0.001 &  0.002 & -325.76 & -0.001 & -0.015 & 2798.64 \\
$\beta_2$                         &  0.991 &  0.356 & -64.09  &  0.749 &  0.512 & -31.68 \\
$\beta_3$                         &  1.946 &  1.492 & -23.33  &  1.782 &  1.892 &  6.18 \\
$\beta_4$                         &  1.547 &  1.545 &  -0.14  &  0.433 &  0.454 &  4.86 \\
$\beta_5$                         &  0.883 &  0.617 & -30.15  &  0.754 &  0.717 & -4.96 \\
$\beta_6$                         &  1.214 &  0.604 & -50.23  &  0.664 &  0.513 & -22.74 \\
$\beta_7$                         &  0.673 &  0.428 & -36.33  &  0.326 &  0.319 & -2.01 \\
$\beta_8$                         &  0.516 &  0.060 & -88.44  &  0.191 & -0.138 & -171.98 \\
\bottomrule
\end{tabular}
\end{table}

\subsection{Comparison of Regression Coefficients for HIV Models}

Table~\ref{tab:coeficientes_vih} displays the estimated coefficients for the logistic regression model for HIV across four approaches: the standard model (STD), the Liu model (simultaneous adjustment of misclassification parameters), the Bayesian model without internal correction (BC), and the Bayesian model with misclassification correction (BEC). Relative percentage changes concerning the standard model are also included.

The intercept showed a reduction of 20.73\% in the Liu model and 12.93\% in the Bayesian model with correction, suggesting differences in the estimation of the baseline prevalence level when diagnostic error is accounted for.

Overall, substantial differences were observed in the magnitude and direction of some coefficients depending on the adjustment method used. The Liu model showed notable reductions compared to the standard model across most coefficients, with a change of up to -88.44\% for the coefficient associated with the sex worker population ($\beta_8$) and -64.09\% for male sex ($\beta_2$), indicating significant adjustment when simultaneously correcting for test sensitivity and specificity.

In contrast, the Bayesian model with misclassification correction (BEC) adjusted the magnitude and direction of specific coefficients. The most striking case is $\beta_8$, which changed from a positive value in the standard model (0.516) to negative (-0.138) in BEC, representing a relative change of approximately 1.7 times compared to the BC model.

Another relevant finding is observed in the coefficient for age group ($\beta_1$), which changed from -0.001 in the standard model to -0.015 in BEC, representing a relative increase of approximately 3 times compared to its value in the BC model. Although its magnitude remains small, this change reflects the sensitivity of the age effect to how misclassification is modeled.

The covariate associated with syphilis reactivity ($\beta_3$) maintained a positive direction in all models, with a slight increase in BEC (1.892) compared to STD (1.946), suggesting the stability of this effect despite adjustment for measurement error.

\begin{table}
\centering
\caption{Comparison of standard errors for HIV model coefficients between Liu and Bayesian model with correction (BEC)}
\label{tab:se_coef_vih}
\begin{tabular}{lccc}
\toprule
\textbf{Coefficient} & \textbf{SE (Liu)} & \textbf{SE (BEC)} & \textbf{Relative Change} \\
\midrule
$\beta_0$                          & 0.694 & 0.237 & 7.539 \\
$\beta_1$                          & 0.004 & 0.006 & -0.639 \\
$\beta_2$                          & 0.229 & 0.228 & 0.008 \\
$\beta_3$                          & 0.311 & 0.168 & 2.424 \\
$\beta_4$                          & 0.847 & 0.553 & 1.345 \\
$\beta_5$                          & 0.230 & 0.200 & 0.327 \\
$\beta_6$                          & 0.415 & 0.397 & 0.089 \\
$\beta_7$                          & 0.387 & 0.402 & -0.073 \\
$\beta_8$                          & 0.172 & 0.311 & -0.695 \\
\bottomrule
\end{tabular}
\end{table}

Table~\ref{tab:se_coef_vih} compares the standard errors (SE) of the estimated coefficients for the HIV model between the Liu approach and the Bayesian model with misclassification correction (BEC). Overall, notable differences were observed in estimator precision depending on the methodological approach.

The BEC model showed a marked reduction in the standard error of the intercept (from 0.694 to 0.237), corresponding to a relative precision gain of more than sevenfold, suggesting increased stability in estimating the corrected baseline prevalence. Similarly, reductions were observed in the standard errors of the covariate associated with syphilis reactivity, with a 2.42-fold relative precision gain compared to the Liu model. A substantial improvement was also observed for the hepatitis B covariate.

In contrast, some covariates—such as age group and sex worker—showed increased standard errors under the BEC model, which may reflect greater uncertainty associated with their effects, possibly due to low frequency or interactions with other variables.

These findings indicate that the BEC model not only adjusts point estimates of the coefficients but also differentially improves their precision, particularly for components most relevant to population-level inference. This reinforces its usefulness in scenarios where diagnostic error may compromise both the validity and precision of traditional models.

\subsection{Comparison of Regression Coefficients for Syphilis Models}

\begin{table}
\centering
\caption{Comparison of regression coefficients for syphilis according to the implemented model}
\label{tab:coeficientes_sifilis}
\begin{tabular}{lcccccc}
\toprule
\textbf{Coefficient} & \textbf{STD} & \textbf{Liu} & \textbf{Liu Change (\%)} & \textbf{BC} & \textbf{BEC} & \textbf{BEC Change (\%)} \\
\midrule
$\beta_0$                         & -4.890 & -3.164 & -35.30  & -4.766 & -5.322 & 11.67 \\
$\beta_1$                         &  0.036 &  0.022 & -38.72  &  0.035 &  0.037 & 5.04 \\
$\beta_2$                         &  0.615 &  0.159 & -74.15  &  0.575 &  0.728 & 26.62 \\
$\beta_3$                         &  1.986 &  1.525 & -23.22  &  1.839 &  2.061 & 12.09 \\
$\beta_4$                         &  2.140 &  1.743 & -18.55  &  0.747 &  0.719 & -3.71 \\
$\beta_5$                         &  0.954 &  0.597 & -37.45  &  0.903 &  1.004 & 11.24 \\
$\beta_6$                         &  1.473 &  0.672 & -54.39  &  1.222 &  1.371 & 12.19 \\
$\beta_7$                         &  1.470 &  0.921 & -37.39  &  1.288 &  1.439 & 11.77 \\
$\beta_8$                         &  1.825 &  0.886 & -51.43  &  1.691 &  1.990 & 17.67 \\
\bottomrule
\end{tabular}
\end{table}

Table~\ref{tab:coeficientes_sifilis} presents the estimated coefficients for the logistic regression model for syphilis under the different models evaluated. Differences in magnitude and direction are observed across approaches, particularly when comparing models corrected for misclassification with the standard model.

The Liu model showed a systematic reduction in most coefficients, with the most notable decreases observed for male sex ($\beta_2$), with a change of -74.15\%, and the sex worker category ($\beta_8$), with a decrease of 51.43\%. These results may reflect an overestimation of associations in the standard model when misclassification is not accounted for.

In contrast, the Bayesian model with misclassification correction (BEC) preserved the positive direction of most effects and even increased their magnitude compared to the standard model. For example, the coefficient for sex workers increased from 1.825 to 1.990, representing a 17.67\% rise, while the effect of male sex increased by 26.62\%. These changes suggest that, once misclassification is corrected, the association between these factors and syphilis infection strengthens—an insight that may be epidemiologically and programmatically significant.

Conversely, the effect of hepatitis B reactivity ($\beta_4$) showed a decrease in the BEC model relative to the standard model, suggesting a potential adjustment toward a more conservative effect. The age group variable ($\beta_1$) maintained a positive and stable effect across all models.

\begin{table}
\centering
\caption{Comparison of standard errors for syphilis model coefficients: Liu model vs. Bayesian model with correction (BEC)}
\label{tab:se_coef_sifilis}
\begin{tabular}{lccc}
\toprule
\textbf{Coefficient} & \textbf{SE (Liu)} & \textbf{SE (BEC)} & \textbf{Relative Change} \\
\midrule
$\beta_0$                         & 0.367 & 0.166 & 3.884 \\
$\beta_1$                         & 0.004 & 0.003 & 0.627 \\
$\beta_2$                         & 0.062 & 0.165 & -0.857 \\
$\beta_3$                         & 0.187 & 0.184 & 0.031 \\
$\beta_4$                         & 0.675 & 0.516 & 0.711 \\
$\beta_5$                         & 0.109 & 0.126 & -0.255 \\
$\beta_6$                         & 0.193 & 0.285 & -0.544 \\
$\beta_7$                         & 0.191 & 0.230 & -0.311 \\
$\beta_8$                         & 0.168 & 0.164 & 0.047 \\
\bottomrule
\end{tabular}
\end{table}

Table~\ref{tab:se_coef_sifilis} shows moderate variations in the precision of the estimators between the two models, with a pattern differing from that observed in the analysis for HIV.

The BEC model demonstrated a substantial improvement in the intercept's precision, with a reduction in the standard error from 0.367 to 0.166, representing a relative precision gain of more than 3.8-fold. This finding suggests that the estimation of the baseline prevalence for syphilis is more stable under a Bayesian approach with misclassification correction.

However, for several covariates—such as sex, LGTBIQ population type, and other population groups—an increase in the standard error was observed under the BEC model. For instance, the SE for sex increased from 0.062 to 0.165, which may reflect more significant uncertainty in estimating this covariate's effect when diagnostic uncertainty is explicitly incorporated. A similar trend was noted for the population group categories, with relative decreases in precision ranging from -25\% to -54\%.

In contrast, variables such as age, hepatitis B reactivity, and sex worker status showed comparable or slightly reduced standard errors under the Bayesian model, suggesting that the estimates' precision was maintained or even improved.

Overall, these results indicate that the Bayesian model with correction adjusts point estimates and alters precision in a covariate-specific manner. The explicit correction of misclassification errors enables a more robust estimation of the intercept (and thus of the adjusted prevalence). However, it may come with trade-offs in the precision of specific individual predictors.

\section{Discussion}

Several studies have demonstrated that estimating disease prevalence using imperfect diagnostic tests can lead to substantial bias if sensitivity and specificity are not adequately accounted for. The need for such correction has been widely documented in the literature, particularly in settings where prevalence varies across populations. In this regard, Li and Fine \cite{Fine} point out that sensitivity—traditionally considered an intrinsic property of the test—is a significant factor influencing prevalence estimates due to factors such as changes in the severity spectrum of cases or variations in the clinical context of diagnosis. Consequently, adjusting for sensitivity and specificity becomes methodologically relevant and necessary to ensure valid and comparable prevalence estimates across studies or heterogeneous populations.

Accurate estimation of prevalence adjusted for individual characteristics becomes especially relevant when the need is not only to obtain an overall population average but also to generate stratified or subgroup-specific estimates defined by demographic or epidemiological covariates. In this context, logistic regression models emerge as an optimal and flexible tool to address this need. Vansteelandt et al. \cite{Vans} emphasize that when information on individual-level covariates is available, the extension of classical estimators through generalized linear models allows for the direct inclusion of these factors in the analysis, thereby improving the precision and validity of prevalence estimates. This approach is not only adaptable to different sampling design configurations. However, it is also advantageous in terms of statistical efficiency and cost-effectiveness, as it makes more effective use of the collected information—especially when combined with methodological proposals that expand its scope for estimation purposes~\cite{villalobos}.

This study compared different methodological approaches for estimating prevalence and associations in the presence of misclassification in the outcome, applied to sexually transmitted infections with low and moderate prevalence. Standard logistic regression models were evaluated, the model proposed by Liu et al. \cite{Liu}, and two Bayesian approaches—with and without explicit correction for misclassification \cite{valle}.

It is important to note that although relevant changes were observed in the coefficients of some covariates across the different models, the magnitude and direction of these changes should be interpreted in light of the underlying distribution of variables in the study population. For instance, coefficients with low absolute values in the standard model may exhibit significant relative variations in the corrected models without necessarily implying a substantive change in their effect.

In contrast, the intercept ($\beta_0$) showed consistent changes across models, suggesting that the primary corrections are reflected in the baseline component of the model—that is, in the prevalence estimation when all covariates are set to zero. In this context, the corrected intercept can be interpreted as approximating the baseline prevalence adjusted for misclassification error, providing a more stable and robust outcome measure without additional risk factors.

These findings reinforce the notion that correcting for misclassification affects relative associations and the point estimate of the baseline prevalence. In particular, it was observed that adjusted estimates differ systematically from unadjusted ones. While it cannot be asserted with certainty that these corrections necessarily yield values closer to the true prevalence, it is reasonable to consider that they represent a significant methodological improvement. In this sense, corrected models may offer estimates more consistent with epidemiological assumptions and less biased due to misclassification error, which is especially relevant in population-based studies and epidemiological surveillance systems.

The results revealed substantial differences in adjusted prevalence estimates and regression coefficients depending on the approach used. In particular, the Bayesian model with misclassification correction (BEC) demonstrated a favorable balance between precision and bias correction, producing intermediate estimates between the standard model and the Liu model while yielding the narrowest confidence intervals. This finding was consistent for both HIV and syphilis.

However, methodological and practical implications must be carefully considered when selecting the statistical approach. The model proposed by Liu et al. \cite{Liu}, especially in its more general formulation that simultaneously estimates the false positive (FP) and false negative (FN) rates, incorporates a substantially more significant number of parameters than standard logistic regression. This increased complexity translates into higher sample size requirements to achieve estimation stability and to ensure convergence of the Fisher scoring algorithm. According to the simulations presented in their work, a minimum sample size of approximately $5{,}000$ observations is required to obtain reliable parameter estimates, including the misclassification rates. This requirement becomes even more critical when the outcome prevalence is low, as the limited number of informative events may severely constrain the model's ability to estimate the additional parameters associated with classification error accurately.

In this study, despite having a considerably large sample size ($n = 11452$), challenges in terms of precision were observed, expressed as wide confidence intervals and high standard errors for several coefficients in the Liu model. This phenomenon was particularly evident in covariates associated with very low-prevalence conditions, such as HIV positivity, where the effective number of cases was relatively limited (1.4\% of the total sample). These findings suggest that even in settings with large sample sizes, the model's performance may be compromised by the low occurrence of the event, thereby limiting its practical applicability in studies of rare infectious diseases.

Additionally, the Liu model exhibited convergence issues in scenarios where the data did not include both misclassification errors. This aspect, linked to the structural assumptions of the error model, may compromise the validity of the approach. In contrast, Bayesian models allow for the incorporation of prior information on sensitivity and specificity, improving estimation stability in smaller samples and enabling a probabilistic characterization of uncertainty. In this study, that advantage was reflected in substantial improvements in the precision of the intercept—directly related to the corrected baseline prevalence—as well as in key coefficients.

Concerning Bayesian modeling, an important methodological advantage lies in incorporating prior information on the regression coefficients, allowing for the regularization of estimates and reducing variability associated with small samples. In this regard, Newman et al. \cite{newman} proposed the formulation of informative prior distributions based on the variance of standardized coefficients and the number of covariates included in the model—a methodology implemented in this study. This approach is beneficial in contexts where classical estimation yields unstable coefficients and extensive standard errors.

However, the use of non-informative priors also holds practical applicability, as demonstrated by Gordóvil-Merino et al. \cite{Merino} through a simulation study. They employed weakly informative priors, providing excellent stability in estimating parameters in logistic regression models applied to small samples. Although the authors acknowledged that this approach does not fully resolve issues arising from asymmetric distributions, they emphasized that including prior knowledge—even minimal—helps mitigate extreme inferences and improve model regularization. It is crucial in studies with small numbers of observations or rare events. Thus, the Bayesian approach emerges as a methodologically sound and flexible alternative for situations where the assumptions of classical models are compromised.

Our findings support using models that explicitly account for misclassification in prevalence and association studies based on imperfect diagnostic tests. While the Liu model represents an improvement over standard logistic regression by addressing systematic bias, its practical and theoretical limitations make it less suitable in specific contexts. In contrast, Bayesian models with misclassification correction—such as the one evaluated in this study—stand out as a robust and flexible alternative, particularly in low-prevalence settings, when diagnostic errors are known or partially known, and when precise estimates are needed to support public health decision-making.

\bibliographystyle{unsrt}  


\bibliography{references}

\end{document}